\documentstyle[12pt,a4]{article}

\begin{document}
\title{\LARGE\bf ON SPECTRAL LAWS OF \\
2D--TURBULENCE IN \\  SHELL MODELS}
\author{\large\bf Peter Frick $^{1,3}$ \& Erik Aurell $^{2,3}$}
\maketitle
\bigskip
\begin{tabular}{ll}
$^1$ & Institute of Continuous Media Mechanics,\\
& Russian Academy of Sciences,\\
& 1, Acad.~Korolyov Str., 614061, Perm, RUSSIA\\\\
$^2$ & Dept.~of Mathematics, University of Stockholm\\
& P.O. Box 6701, S--113 85 Stockholm, SWEDEN\\\\
$^3$ & Center for Parallel Computers,\\
& Royal Institute of Technology,\\
& S--100 44 Stockholm, SWEDEN
\end{tabular}

\bigskip

\begin{abstract}
We consider a class of shell models of 2D-turbulence. They
conserve inertially
the analogues of energy and enstrophy,
two quadratic forms in the shell amplitudes.
The coefficients in the inertial interactions
have previously been derived from the 2D Navier-Stokes' equations
under a series of assumptions (Frick 1983).
Inertially conserving two quadratic integrals leads to two
spectral ranges.
We study in detail the one
characterized by a forward cascade of enstrophy and spectrum
close to Kraichnan's $k^{-3}$--law.
In an inertial range over
more than 15 octaves, the spectrum falls off as $k^{-3.05\pm 0.01}$,
with the same slope in all models.
We identify a ``spurious'' intermittency effect, in that the energy spectrum
over a rather wide interval adjoing the viscous cut-off, is well approximated
by a power-law with fall-off significantly steeper than $k^{-3}$.
Assuming Kraichnan's law, we evaluate the prefactor in the
power law to be of the same order of magnitude as in true 2D-turbulence.
\end{abstract}
\medskip
\noindent

\newpage
\section{Introduction}
\label{s:introduction}
The basic idea of shell models of turbulence is to describe fluctuations of
velocity
(or vorticity, or temperature, etc.) in an octave of wave numbers
$2^n<|k|<2^{n+1}$
by one ``collective'' variable $U_n$. This range of wave numbers is called
a shell (the $n$'th shell), and the variable $U_n$ is called a shell
variable.
In this way one obtains a
dynamical system of limited number of degrees of freedom,
describing a cascade process of energy (or enstrophy, or temperature)
in a large interval of wave numbers. The general form of shell equations
can be written for real or complex shell variables as
\begin{equation}
d_t U_n = \sum\limits_{m,l} i X_{nml} U_m^* U_l^* - K_n U_n + f_n
\label{shellZ}
\end{equation}
where $^*$ stands for complex conjugation, $f_n$ is the external force and
$-K_nU_n$ models
the action of the viscous force. The energy in the $n$'th shell is
taken to be proportional to $|U_n|^2$.

Recently shell models have attracted new interest. The Gledzer-Yamada-Ohkitani
model\cite{Gledzer, Yamada} was extensively investigated
by several groups\cite{PisFr, Jensen}, not only numerically,
but also analytically\cite{BenParBif}.
This is a model for 3D-turbulence, written
for real\cite{Gledzer} or complex\cite{Yamada} shell
variables, and takes into account only nonlinear interactions between
neighbouring shells:
\begin{equation}
d_t U_n = i (a_n U_{n+1}^* U_{n+2}^* + b_n U_{n-1}^* U_{n+1}^* + c_n U_{n-1}^*
U_{n-2}^*) -
Re^{-1}k_n^2 U_n + f_n
\label{Yam-Okh}
\end{equation}
Here $Re$ is the Reynolds number, $k_n$ a typical wave number in the $n$'th
shell, and the choice of three coefficients, $a_n=k_n$, $b_n=-k_{n-1}/2$ and
$c_n=-k_{n-2}/2$, ensures
energy conservation in limit $Re\to\infty$, $f_n\to 0$.
We will only consider the case when $k_n=2^n$.

There are two sharp qualitative differences between shell models
generally considered, and the Navier-Stokes' equation: first shell
variables cannot describe spatial fluctuations within one shell, because
they have only one variable per shell; and, secondly, equations
like (\ref{Yam-Okh})
only include local interactions in $k$-space.

The first shortcoming is the inevitable price to pay
for the simplicity and low dimensionality of the system.
One can still obtain interesting models by considering a hierarchical system
in which the number of variables in the shell $n$ grows exponentially
with $n$.
Such a model, for 2D-turbulence, was investigated in \cite{AuFrSh}.
Here however
the number of degrees of
freedom again grows very large, and one cannot consider much larger
intervals of wave number, than in full-scale simulations of Navier-Stokes'
equation.

The second shortcoming can be remediated already within a shell
model. For 2D-turbulence, this was done in an earlier paper by one
of authors\cite{Fr1}. We shall consider this class
of models, only here rewritten for complex shell variables.

\section{Shell model of 2D-turbulence}
\label{s:hierarchical}
The main particularity of 2D Navier-Stokes equations is the
appearence of an infinite number of additional
integrals of motion of the inviscid equations.
The most striking one is enstrophy, the integrated squared vorticity.
To reproduce the general properties of 2D-turbulence, it is essential
to at least conserve energy and enstrophy in the infinite Reynolds number
limit.

In \cite{Fr1} shell model equations of 2D-turbulence
were obtained in the form
\begin{equation}
\begin{array}{l}
d_t U_n = \sum\limits_{j=1}^{J} \Bigl \lbrace T_{n,n-j,n+1}U_{n-j}U_{n+1} +
T_{n,n+j,n+j+1}U_{n+j}U_{n+j+1} + \\ \\
\qquad\qquad T_{n,n-j-1,n-1}U_{n-j-1}U_{n-1} \Bigl \rbrace + K_n U_n + f_n
\end{array}
\label{casFr}
\end{equation}
If $J=1$, only local interactions are kept,
and one returns to the form of (\ref{Yam-Okh}).

The analogues of energy and enstrophy are the quadratic
forms $\sum_n |U_n|^2$ and $\sum_n |k_nU_n|^2$.
Demanding energy and enstrophy
conservation gives two relations between the three
elements $T_{nml}$ inside the
brackets in (\ref{casFr}), and we can rewrite
the equations as:
\begin{equation}
\begin{array}{l}
d_t U_n = i2^n \sum\limits_{j=1}^{J} T_j \Bigl \lbrace
{{3\cdot2^j}\over{4-2^{-2j}}}
U_{n+j}^*U_{n+j+1}^* - U_{n-j}^*U_{n+1}^* + 
{{2^{2j}-1}\over{2^{2j+3}-2}} U_{n-j-1}^*U_{n-1}^* \Bigl \rbrace + K_n U_n +
f_n
\end{array}
\label{casZ}
\end{equation}
The coefficients $T_{j}$ were computed
from a projection
of the 2D Navier-Stokes' equations on a hierarchical function
basis in \cite{Fr1}, constructed from basis functions that have
support in one octave in wave-number\footnote{This is to begin
with a simple wavelet
basis, compact in Fourier space but rather slowly falling off in
real space. In the successive derivations of the model, the orthogonality
property of the wavelet basis is lost, and we therefore rather
refer to it as a hierarchical basis.}.
The normalisation of shell energy and shell enstrophy then come
out as $E_n = {{\log 2}\over{4\pi}}|U_n|^2$, and
$\Omega_n ={{3\pi}\over 8}|2^nU_n|^2$.
We refer to \cite{Fr1} and \cite{AuFrSh} for a discussion of this procedure,
and here just list the first values of $T_{j}$:
$T_1=0.269$, $T_2=0.0795$, $T_3=0.0257$, $T_4=0.0088$ and $T_5=0.00295$;
roughly speaking, the interaction strength decreases by a factor
about three when distance between levels increases by one.

We are in this paper mainly
interested in the qualitative properties of the model,
which (seemingly) depend only on the conservation of two quadratic integrals,
and not on the numerical values of the interaction coefficients.
We have therefore performed numerical experiments, where we keep just one
interaction coefficient at a time, setting the other ones to zero,
and compared the results to keeping all of them.
We have though also evaluated the
Kolmogorov--Kraichnan constant $C_{\omega}$, under
the (not quite true) assumption that the spectrum in
the inertial range obeys:
\begin{equation}
E(K)\sim C_{\omega}\epsilon_{\omega}^{2\over 3}
k^{-3}\qquad\quad\hbox{(Kraichnan's $k^{-3}$-law)}
\label{law}
\end{equation}
where $\epsilon_{\omega}$ is the mean enstrophy dissipation.
We here find order of magnitude agreement with reported values
in the literature, partially justifying the procedure
used to obtain the coefficients $T_j$.

We have also for completeness compared
the results using real or complex shell
variables. The main difference is that the complex model can
fluctuate in phase while keeping a more steady amplitude, while the
real model must go through zero amplitude to change sign. It may
therefore be argued that the complex model is qualitatively more
similar to true turbulence, than is the real model.
In  fig.\ref{f:ens_z} we show
the time evolutions of total enstrophy $\Omega$ and the rate
of enstrophy dissipation $\epsilon_\omega$ in the complex model.
In the real model
the value of total enstrophy fluctuates more, but the peaks
of enstrophy dissipations rate are less sharp.
\begin{figure}
\input{fig1.tex}
\caption{Time evolution of the enstrophy $\Omega$ and the rate of enstrophy
dissipation $\epsilon_{\omega}$.}
\label{f:ens_z}
\end{figure}
When we compute long averages over time, we however find
that the real and complex models give the same result, only
the complex model needs less time to approach a statistically
stationary state.

\section{Numerical results }
\label{s:numerical}
In our numerical experiments the external force
is random in time, with rms-amplitude order unity,
and acts on the first and second (or fourth and fifth) shells.
Energy is transported by the inverse cascade from where
the force acts down towards the low $n$ shells,
where we have to remove it to obtain a stationary regime.
We implemented this condition in the standard
way, by adding to the equations a linear friction term,
acting in the zeroth shell ($f_0 = -F_0 U_0$). The number of shells
was 31 ($0 \leq n \leq 30$). All the figures give the data for Reynolds
number $Re=6.3\cdot 10^{18}$.

To integrate the equations we used a fourth-order Runge-Kutta integrator
with time step 0.1. We remark that in 2D--turbulence, and in our shell
models, the characteristic times are approximately the same on all spatial
scales in the inertial range.
The accuracy of the integration method was checked by repeating
the integrations using smaller step-size.
Runs up to $20\cdot 10^6$ time-steps were made.
Unless where indicated, all figures show data sufficiently converged
so that the statistical scatter is not visible, and simulations of the
model using the full set of interaction coefficients $T_j$.

\begin{figure}
\input{fig2.tex}
\caption{ Time-averaged shell energy  distribution. $Re=6.3\cdot 10^{18}$.}
\label{f:speclar}
\end{figure}
\begin{figure}
\input{fig3.tex}
\caption{The increment enstrophy spectrum,
$\alpha_n = {1\over{\log 2}} \log{{\Omega_{n+1}}\over{\Omega_n}}$.
$20\cdot 10^6$ time-steps. Fluctuations are still, but barely, visible in this
statistics after this time.}
\label{f:slop1}
\end{figure}

In fig.\ref{f:speclar} we show the shell energy distribution.
The two straight lines drawn seem to indicate
that the spectrum consists of two parts: one from where
the force acts
up to $n$ about 20, with slope $E(k) \sim k^{-3.05 \pm 0.01}$,
and one from $n$ around 20 to about 26,
with appearent slope $E(k) \sim k^{-3.33}$.
For higher $n$ the
spectrum is visually curved in a log--log plot (see fig. \ref{f:speclar};
we are evidently in the dissipative range.

One can display the data in another way, that clearly shows
that there is in fact only one power law (with corrections towards the viscous
end).
Consider (fig.\ref{f:slop1}, where we have plotted the logarithms of the
quotients between enstrophies of two succesive shells. Let us recall
that if the energy spectrum goes as $k^{-3}$, then the enstrophy
spectrum goes as $k^{-1}$, and the enstrophy content of a
shell is constant. If the shell enstrophy decreases with shell
number as $2^{-n\alpha}$, then the base two logarithms of two successive
quotients tend to $-\alpha$.
It is clear from  (fig.\ref{f:slop1}) that the slope of the spectrum
goes down smoothly through the second range, and that one cannot
associate a second power-law to it.
One may also fit a power-law behaviour with an expontial
cut-off to the data. Fig (\ref{f:expdata}) shows our best fit of the
logarithms of shell enstrophies.
Clearly this fit is reasonably good in the inertial range with $n$
between $5$ and $20$, good in the far dissipation range,
but not very good in the intermediate region.
Using the slope of the linear fit of
fig.\ref{f:speclar}, we can get a better fit in the inertial
range, at the price of having a much worse fit in the
intermediate range.
We therefore conclude that the corrections to the power-law
in fig.\ref{f:speclar} are not exponential, and not a second
power-law, but something in between\footnote{An
ansatz following Frisch \& Vegassola\cite{FV} of an intermediate dissipation
range
with pseudo-algebraic fall-off can indeed fit the data very well.
As the number of parameters of this fit is rather high (at least three more
than in the exponential fit) we are not sure if this
is significant information.}.
\begin{figure}
\input{fig4.tex}
\caption{Least square fit of shell enstrophies in logarithmic scale
to a parametrization $A + (\alpha ln 2)\cdot n - \exp(\beta(n-n^*))$.
$A=7.30$, $\alpha=0.0742$, $\beta=0.569$ and $n^* = 26.2$.}
\label{f:expdata}
\end{figure}

We note that there have been earlier discussions of ``spurious''
intermittency corrections\cite{AFLV},
and that we here have one of the mechanisms reported there:
a correction to a power law which is similar to a weaker
power-law, is easily mistaken for a power-law with another slope.

\begin{figure}
\input{fig5.tex}
\caption{Logarithm of the flatness factors versus the shell number.
$10^6$ time steps.}
\label{f:flatl}
\end{figure}

The intermittency corrections to the $k^{-3}$-law (\ref{law})
show up in the
behaviour of the structure functions
$S_q(l)$, definded as the space averages of $q$-th moment of
fluctuations of scale $l$.
In the shell model the structure functions
of shell enstrophy are computed as the time average values of
corresponding variable in power $q$.
\begin{equation}
S_q = <|U_n|^{2q}>
\label{struct}
\end{equation}

If (\ref{law}) is true, in the inertial range
\begin{equation}
S_q(r) \sim l^{\eta_q}
\label{etadef}
\end{equation}
and  the exponents follow the simple law
$\eta_q \sim q h $  with $h = 1$.

The traditional measure which estimates the deviation
of a probability distribution
from a Gaussian distribution, is the flatness factor, computed as the ratio of
fourth
moment to the square of second moment, minus three (the flatness of a Gaussian
distribution is then zero). In our case we compute the flatness factor
for the enstrophy on shell $n$ as
\begin{equation}
\gamma_n = {{< (|U_n|^2)^4>} \over { (<(|U_n|^2>)^2}} -3.
\end{equation}
If the flatness factor grows exponentially with shell number, then
the growth rate is connected with the exponents in (\ref{etadef})
${\it ln} \gamma_n \sim  n {\it ln}2(\eta_4-2\eta_2) = \lambda n$.
A non-zero $\lambda$
shows the deviation from $\eta_q \sim q h $.

The flatness factors of the distribution of shell enstrophy is shown in
fig.\ref{f:flatl}.
One can here again distinguish the same two sub-intervals of the
inertial range as in the energy spectra,
only now more pronounced.
The flatness factor increases very slowly in the first interval,
approximately with slope $\lambda = 0.024$ in the range $n=[5,20]$.
This is quite consistent with a small correction to the
$k^{-3}$-spectrum.
In the range $n=[20,26]$ a fit gives a value of $\lambda = 0.13$,
i.e. five times higher.
The flatness factor converges much more slowly than the energy
spectrum, and we still have significant scatter in fig.\ref{f:flatl}.
We do not doubt however that the appearent second power law in the
flatness factor, is as spurious as was the one in the energy spectrum.

We emphasize that this appearent strong difference
in intermittency
corrections in different ranges, and in general
the spurious power-law range, may in 2D be a property
of shell-models only.
Shell models are a kind of mean-field theories, and
it may be to expect too much that they would reproduce
not only the spectrum, but also fluctuations:
temporal intermittency in the shell model can be
quite different from
spatial intermittency.

\begin{figure}
\input{fig6.tex}
\caption{ Time-averaged level enstrophy, keeping only one interaction
(dots and bullets), or keeping all of them (points).}
\label{f:differJ}
\end{figure}

Let us now turn to the question of different shell models (\ref{casZ})
and their spectrum.
We have taken the extreme limit and kept in the simulations
displayed in fig.(\ref{f:differJ}) only one type of
interaction in each
(a run where all are kept is shown for reference).
Although these models give the same spectral slope (\ref{f:differJ}),
they are in two respects quite different: both in values of their
interaction coefficients; and in the choice of interacting triads.
We therefore conjecture, that the spectral law of 2D-turbulence in
shell models, albeit slightly different from the $k^{-3}$ prediction,
nevertheless holds for a wide class of models, that inertially
conserve two quadratic forms in the shell amplitudes, i.e. is universal.

The curves in (\ref{f:differJ}) with only one type
of interaction are shifted relative to
the reference run.
This may be understood as follows: the full model
transports enstrophy more effectively
through the spectrum. In the inertial range, the mean value of shell
enstrophies is therefore lower. In the dissipation range
the mean value of shell enstrophies is instead higher,
because these shells are then driven more intensively
by the
inertial range.

Finally we can try to estimate the Kolmogorov--Kraichnan constant
using (\ref{law}). As in fact the spectrum does not fall off
precisely as
$k^{-3}$ in the inertial range, we get estimates of $C_{\omega}$
that change slowly with $k$. Using our normalisation of the shell energy
and the data in fig.\ref{f:speclar}, we estimate $C_{\omega}$ to be
6 and 10 at, respectively, the top ($n=5$), and the bottom ($n=20$),
of
the inertial range. Most of the estimates from numerical studies
of the full Navier-Stokes' equations are in the range 4 to 10.
Our shell model therefore seems to give an order of magnitude
correct estimate of the overall energy content of 2D--turbulence.

{\bf Acknowledgements:} This work was supported by the Wenner-Gren Center
Foundation~(P.F.),
and by the Swedish Natural Science Research Council under contract
S--FO~1778-302~(E.A). We thank the Center for Parallel Computers for
hospitality.

\end{document}